\documentstyle[amssymb,twocolumn,floats,prl,aps,epsfig]{revtex}

\begin{document}
\draft
\wideabs{
\title{Beyond Eliashberg superconductivity in MgB$_2$: anharmonicity,
two-phonon scattering, and multiple gaps}
\author{Amy Y. Liu,$^{1,2}$  I. I. Mazin$^{2},$ and Jens Kortus,$^{1,2,3}$}
\address{$^1$ Department of Physics, Georgetown University, Washington, DC
20057}
\address{$^{2}$ Center for Computational Materials Science, Code 6390, Naval
Research Laboratory, Washington, DC 20375}
\address{$^{3}$ MPI f{\"u}r Festk{\"o}rperforschung, Stuttgart, Germany}
\date{\today }
\maketitle

\begin{abstract}
Density-functional calculations of the phonon spectrum and electron-phonon
coupling in MgB$_2$ are presented. The E$_{2g}$ phonons, which involve
in-plane B displacements, couple strongly to the $p_{x,y}$
electronic bands. The isotropic electron-phonon coupling  constant is
calculated to be about 0.8.  Allowing for different order parameters in 
different bands,
the superconducting $\lambda$ in the clean limit is calculated
to be significantly larger. The $E_{2g}$ phonons are strongly anharmonic,
and the non-linear contribution to the coupling between the $E_{2g}$ modes
and the p$_{x,y}$ bands is significant.
\end{abstract}

}
The recent discovery of superconductivity near 40 K in MgB$_{2}$ has
generated much interest in the properties of this simple intermetallic
compound \cite{magamatsu01}. A significant B isotope
effect strongly suggests  phonon-mediated pairing \cite{budko01}. 
To explain the large $T_c$, an electron-phonon coupling (EPC) 
constant of $\lambda \approx 1$ is needed. 
Yet estimates of the  coupling  strength based on the
latest measurements of the low-temperature specific heat \cite{specificheat},
combined with the density of states (DOS) from density-functional calculations
\cite{kortus01}, yield $\lambda \approx $ 0.6-0.7.
Further, the measured temperature dependence of the electrical resistivity 
\cite{rho} is consistent with $\lambda _{tr}\lesssim 0.6$. 
First-principles calculations of the EPC give 
$\lambda \approx 0.7-0.9$ \cite{kortus01,kong01,bohnen01}.
Clearly there is a problem in reconciling all these numbers. Another puzzle
involves tunneling measurements of the gap. Values of $2\Delta /k_{B}T_{c}$
ranging from 1.2 to 4 have been reported. The values below the BCS weak
coupling limit of 3.5 have been attributed to surface effects, but the
best-quality spectra\cite{rubio} show a very clean gap with $2\Delta
/k_{B}T_{c}=1.25.$  Sharvin contact measurements\cite{sharvin} reveal a gap
at 4.3 meV ($2\Delta /k_{B}T_{c}=2.6)$, and additional structures at
$2\Delta /k_{B}T_{c}=1.5$ and 3, raising the possibility of multiple
gaps. Careful analysis of the temperature and magnetic field dependence of
the specific heat suggests anisotropic or multiple gap structure as well
\cite{specificheat}. 
Thus, even if
superconductivity in MgB$_{2}$ is phonon-mediated, it is likely
that an analysis beyond the simple isotropic Eliashberg model is needed.

The MgB$_{2}$ lattice consists of two parallel systems of flat layers. One
layer contains B atoms in a honeycomb lattice, the other Mg
atoms in a triangular lattice halfway between the B layers.
First-principles
calculations\cite{kortus01}
 find that the electronic states near the Fermi level are
primarily B in character and the Fermi surface (FS) comprises four sheets: 
 two nearly cylindrical hole sheets about the $\Gamma$-A line arising
from quasi-2D $p_{x,y}$ B bands, and two tubular networks arising from 3D
$p_{z}$ bonding and antibonding bands \cite{kortus01}. The difference in 
character between the sheets raises the possibility that each has a distinct
gap that could be observed in the clean limit. Such
interband anisotropy enhances the effective EPC constant relevant to
superconductivity and decreases the coupling  constant
for transport, compared to the  average
values \cite{suhl59,butler76,mazin93}.
This could explain the discrepant values of $\lambda$ deduced from 
different types of experiments. 

In this paper, we report first-principles calculations of the
EPC in MgB$_{2}$. The coupling 
constant is decomposed into contributions from the four different bands
crossing the Fermi level, allowing for an analysis of the effects of
interband anisotropy on the superconducting $T_{c}$ and gap structure. The
strongest coupling arises from the $E_{2g}$ phonon modes along
the $\Gamma$ to A line, which strongly interact with the quasi-2D electronic
states. This phonon mode is calculated to be highly anharmonic, and it also
has significant nonlinear contributions to the EPC\cite{yld}.

Harmonic phonon frequencies and linear EPC parameters
were calculated using the  linear-response method
within the  local density approximation 
\cite{quong92}. Norm-conserving
pseudopotentials \cite{troullier91} were used, with 
a plane-wave cutoff of 50 Ry. 
We used the experimental crystal structure, with  
$a=3.08$~{\AA}~ and $c/a=1.14$
\cite{muetterties67}. The electronic states were sampled on grids of up to
$24^3$ {\bf k}-points in the full Brillouin zone, and the dynamical
matrix was calculated on a grid of $8^3$  phonon wavevectors {\bf q} \cite{footnote1}.

\begin{figure}[tbp]
\centerline{
\epsfig{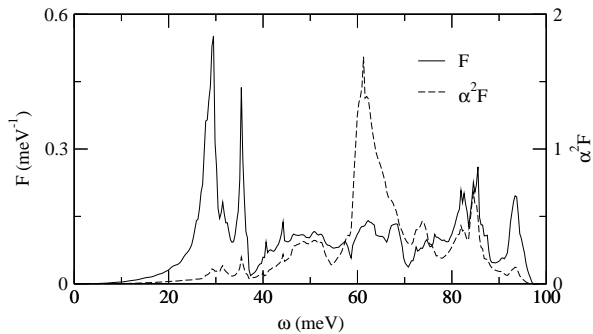}}
\caption{ Phonon density of states and Eliashberg function. }
\label{fig1}
\end{figure}

The calculated phonon density of states $F(\omega)$ and
Eliashberg function $\alpha^2F(\omega)$ are plotted in Fig.\ 
\ref{fig1}. The results are similar to those reported in 
Refs. \cite{kong01} and \cite{bohnen01}. 
All of these calculations give a slightly softer phonon spectrum  
than what is observed in neutron experiments \cite{osborn01}. 
While $F$ and $\alpha ^{2}F$ are similar in
shape in many materials, they are strikingly different in
MgB$_{2}$. In particular, 
$\alpha ^{2}F$ has a pronounced peak in the range of 60 to 70 meV
arising from dispersive optic modes that do not give rise to large structures
in $F$. Correspondingly, the average phonon
frequency $\omega _{ave}=55.3$ meV is less than
 the logarithmically averaged frequency  
$\omega _{\ln}=\exp [{\lambda }^{-1}\int \ln {\omega }\alpha ^{2}F(\omega
)\omega ^{-1}d\omega ]=56.2$ meV, despite the fact that logarithmic averaging
preferentially weights lower frequencies.  The 
isotropic EPC constant, which determines $T_c$ in the dirty limit,
 $\lambda _{sc}^{0}=2\int{\omega^{-1}\alpha ^{2}F(\omega)d\omega}$ 
is found to be 0.77, in reasonable
agreement with other calculations \cite{kortus01,kong01,bohnen01}. 

The peak in $\alpha ^{2}F$  between 60 and 70 meV arises from the 
$E_{2g}$ phonon modes with {\bf q} along the $\Gamma$-A line. This
Raman active phonon mode, doubly degenerate at $\Gamma$, involves in-plane,
hexagon-distorting displacements of the B atoms. In fact, by symmetry, this
is the only mode at $\Gamma $ that has a linear EPC. Going away from
the $\Gamma$-A line the EPC drops sharply when the phonon wave vector 
{\bf q} becomes larger than the diameter of the 2D Fermi surface; 
at the same time the frequency increases by roughly 30\%. 
This indicates that the 
reason why this B-B bond-stretching mode is not the highest-frequency
mode at $\Gamma $ is because of softening due to EPC. 
However, this softening should weaken in
the superconducting state, since some of the screening electrons form
Cooper pairs and are removed  from the Fermi sea. Such hardening of
optical phonons below $T_{c}$ was first discussed by Zeyher and Zwicknagl %
\cite{zeyher90}. The overall scale of the relative hardening, $\Delta \omega
/\omega ,$ is set by a specific EPC constant, 
$\lambda _{ZZ}={{2}{\omega^{-1} }}\sum_{{\bf k}i}|g_{k,k}|^{2}\delta
(\epsilon _{{\bf k}i}),
$
where $g$ is the EPC matrix element.  (The 
Fermi level is set to zero.) In the BCS limit, $\Delta
\omega /\omega $ is a known analytical function\cite{osv}
 of $\omega$. We calculate
$\lambda_{ZZ}=0.6$, for the $E_{2g}$ mode. Taking $\Delta \sim 5$ meV
we predict about a 12\% hardening of this
mode below $T_{c}$. This shift should be easily observable in Raman or
neutron experiments.

Since 
the 2D FSs are calculated to play an important
role in 
the EPC, 
we have decomposed the relevant electronic characteristics in terms of the 
four sheets of the FS. We list in Table I the 
partial DOS $N_{i}=\sum_{{\bf k}%
}\delta (\epsilon _{{\bf k}i})$, and plasma frequencies $\omega
_{p,i,\alpha \alpha }^{2}={\frac{8\pi e^{2}}{V}}W_{i}={\frac{8\pi e^{2}}{V}}%
\sum_{{\bf k}}v_{{\bf k}i,\alpha }^{2}\delta (\epsilon _{{\bf k}i})$,
with $i=1(2)$ referring to the light(heavy)-hole 2D sheets 
of the Fermi surface, 
and $i=3(4)$ to the $p_{z}$ bonding (antibonding) sheets.
The EPC   constant was also 
decomposed into contributions from scattering of an electron
from band $i$ to band $j$: 
\begin{eqnarray*}
\lambda _{sc}^{0} &=&\sum_{ij}U_{ij}N_{i}N_{j}/N=\sum_{i}\lambda _{i}N_{i}/N
\\
U_{ij}N_{i}N_{j} &=&2\sum_{{\bf kq}\nu }{\omega^{-1} _{{\bf q}\nu }}
|g_{{\bf k}i,{\bf k+q}j}^{\nu }|^{2}\delta (\epsilon _{%
{\bf k}i})\delta (\epsilon _{{\bf k+q}j}).
\end{eqnarray*}%
Here $\omega _{{\bf q}\nu }$ is the frequency of the corresponding phonon,
and $\lambda _{sc}^{0}$ is the standard (Eliashberg) isotropic coupling
constant. 
Allowing for interband anisotropy of the order parameter (clean limit), the
effective coupling  constant for superconductivity $\lambda _{sc}^{eff}$ is
given by the maximum eigenvalue of the matrix $\Lambda _{ij}=U_{ij}N_{i}$,
which is always larger than $\lambda _{sc}^{0}$. Assuming the same
interaction parameters $U_{ij}$ for transport properties, the lowest order
variational approximation for the Boltzmann equation corresponds to the
transport EPC constant $\lambda _{tr}^{0}=\sum_{i}\lambda _{i}W_{i}/W$. On
the other hand, allowing variational freedom for the different sheets of
the Fermi surface yields an effective transport coupling constant which is
always smaller than $\lambda _{tr}^{0}$. In effect, the different bands
provide parallel channels for conduction, so that when ``scattering-in'' is
neglected, $W/\lambda _{tr}^{eff}=\sum_{i}W_{i}/\lambda _{i}$ \cite{mazin93}.

The calculated interaction parameters $U_{ij}$ are listed in Table II. 
Due to similarities between the two 2D sheets, and
between the two 3D sheets, 
we have simplified the model to allow for two
different order parameters for these two sets of bands.
This gives $U_{AA}=0.47$
Ry, $U_{BB}=0.10$ Ry,  and $U_{AB}=0.08$ Ry, where $A$ and $B$ stand 
for the 2D and 3D bands, respectively. 
Then $\lambda _{A}=1.19$ and $\lambda _{B}=0.45,$
suggesting de Haas-van Alphen mass renormalizations 
of $\sim 2.2$ and $\sim 1.5,$ for the two sets of bands, and specific
heat renormalization of 1.77\cite{sheat}. The resulting anisotropic
effective coupling constant for superconductivity is $\lambda
_{sc}^{eff}=1.01$. Using the Allen-Dynes approximate formula for $T_{c}$ %
\cite{allen75}, we find that to have $T_{c}=40$ K, a Coulomb pseudopotential
of $\mu ^{\ast }\approx 0.13$ is needed. This is more physically plausible
than the $\mu ^{\ast }\approx 0.04$ required when $\lambda _{sc}^{0}$ is
used. For transport, interband anisotropy reduces the in-plane
coupling constant $\lambda _{x,y}$ from 0.70 to 0.57, but has essentially
no effect on the out-of-plane $\lambda _{z}=0.46$ (Table III).  
This is because the
anisotropic formula accounts for the fact that the transport is mostly due
to the 3D bands in any direction, simply because they couple less with
phonons. 
The measured resistivity \cite{rho} can be fit 
remarkably well with the Bloch-Gr\"{u}neisen formula using
the calculated isotropic $\alpha_{tr}^2F$, with the in-plane and
out-of-plane contributions  appropriately averaged for polycrystalline
samples \cite{stroud75}. 
However, with band anisotropy, 
the resistivity is slightly underestimated. 

\begin{figure}[tbp]
\centerline{
\epsfig{file=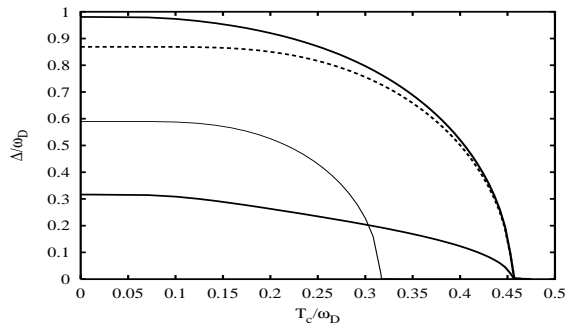,width=0.9\linewidth,clip=true}}
\caption{Superconducting order parameters $\Delta_i$ in the multigap
weak-coupling approximation (solid lines), and in the isotropic (dirty)
BCS limit (thin line). The BCS order parameter corresponding to 
the the same $T_c$ as the multigap model is shown by the dashed line.}
\label{gaps}
\end{figure}

The temperature dependence of the individual gaps $\Delta _{i}$ in the
weak-coupling multigap model
is defined by
$\Delta _{i}=\sum_{j}U_{ij}N_{j}\Delta _{j}\int dE\tanh
(\frac{\sqrt{E^{2}+\Delta_j^{2}}}{2T})/\sqrt{E^{2}+\Delta_j^{2}}.$
As shown in Fig.\ \ref{gaps}, 
the larger 2D gap is calculated to be BCS-like, with
a slightly enhanced $2\Delta /T_{c}$,  while the 3D gap is about three
times smaller in magnitude. Thus, in the clean limit, MgB$_2$
should have two very different order parameters, which in turn should
affect thermodynamic properties in the superconducting
state. Experiments indicate that the coherence length in MgB$_{2}$ is
close to 50~{\AA}. The mean free path corresponding to the residual resistivity
observed in Ref.\ \cite{rho} is more than 1000~{\AA}, so that
$2\pi\xi /l\approx 1/3.$ This is in the reasonably clean regime, 
and it is likely
that the intrinsic resistivity is even smaller. However, stronger defect
scattering should be detrimental to superconductivity: using the Allen-Dynes
formula with the same $\mu ^{\ast }=0.13,$ we get an isotropic $T_{c}=22$ K.
Indeed, irradiation has been found to drastically reduce $T_{c}$%
\cite{irrad}. Some of the experimental manifestations of 
multigap superconductivity would be a reduced and impurity-sensitive
specific-heat jump at $T_{c}$, a deviation of the critical-field temperature
dependence from the Hohenberg-Werthamer formula, a reduction of the
Hebel-Slichter peak in NMR, and a substantial difference between the
in-plane and out-of-plane tunneling spectra. In particular, the latter 
should see only the smaller gap\cite{note3}.

Note that 
$\lambda $ $\sim 1$ is in the intermediate coupling regime. Furthermore,
the multigap scenario suggests particular sensitivity to impurity scattering.
This means one should really solve the anisotropic Eliashberg equations 
with impurity scattering, rather than the weak coupling BCS equations
we used. Thus we do not make any quantitative thermodynamic and 
spectroscopic predictions here. 

We focus now on the $E_{2g}$ phonon
modes, which carry the lion share of the EPC. We have
examined this mode at $\Gamma$ in  detail using frozen-phonon
calculations. The calculations were carried out using a general potential
LAPW code with the same setup as in Ref.\ \cite{kortus01}.
This mode  has substantial anharmonicity.  
A fit of the total energy for B displacements $u$
between $\pm 0.07$ a.u. 
to a fourth-order polynomial ($E_{tot} = \sum a_n u^n$) 
gives $a_2 = 0.434$ Ry/a.u.$^2$, $a_3 = -1.868$ Ry/a.u.$^3$ and 
$a_4 = 14.926$ Ry/a.u.$^4$.   
Anharmonicity increases the
frequency of this mode by about 15\%, 
which should result in an overall reduction
of $\lambda $ by $\sim$10\%, and an increase of $\omega _{\ln }$ by $%
\sim$6\%.

\begin{figure}[tbp]
\centerline{
\epsfig{file=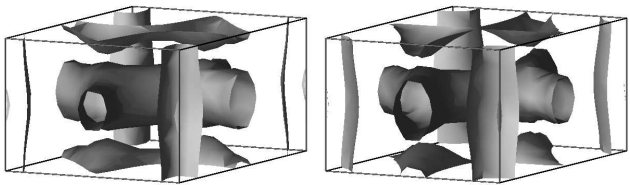,width=1\linewidth,clip=true}}
\caption{Fermi surfaces for two frozen-phonon patterns 
of $E_{2g}$ symmetry with B displacements of $\pm 0.07$ a.u.  
One of the cylindrical sheets has shrunk and is barely visible.
}
\label{fs}
\end{figure}

More interestingly, the $E_{2g}$ modes have a significant
nonlinear coupling with electrons. The linear coupling vertex, $g_{1}$,
corresponding to scattering by a single phonon, is proportional to matrix
elements of $dV/dQ$, where $Q=\sqrt{2M\omega}u,$ while the second-order
coupling, involving exchange of two phonons, is proportional to matrix
elements of $d^{2}V/dQ^{2}$. At $\Gamma$, the Hellman-Feynman theorem
allows the calculation of $g_{1}$ via deformation potentials. This is no
longer the case for $g_{2}$.  One can 
use $d^{2}\epsilon _{{\bf k}}/dQ^{2}$ only as a qualitative estimate
of $\langle d^{2}V/dQ^{2}\rangle$. 
For the cylindrical sheets of
the Fermi surface, $\left\langle 
(d^{2}\epsilon _{{\bf k}}/dQ^{2})^2\right\rangle^{1/2} = 26$ and 20  mRy
as compared to $\left\langle (d\epsilon _{{\bf k}}/dQ)^2 \right\rangle^{1/2} 
= 14$ and 16  mRy. This suggests that nonlinear pairing via {\it
two-phonon} exchange is comparable to  or even larger than the linear coupling%
\cite{isotopenote}. The reason for this anomalous behavior 
lies in the
specifics of the band structure of the 2D $p_{x,y}$ bands. 
In the nearest-neighbor tight-binding approximation it can be described as
\begin{eqnarray*} \epsilon _{{\bf k}}^{2} &=&u_{{\bf k}}\pm v_{{\bf k}} \\
4u_{{\bf k}} &=&(t_{\pi }^{2}+t_{\sigma }^{2})(6+\sum_{i}\cos G_{i})+6t_{\pi
}t_{\sigma }\sum_{i}\cos G_{i} \\ (4v_{{\bf k}})^{2} &=&4(t_{\pi }^{2}-t_
{\sigma }^{2})^{2}(\sum_{i}\cos^{2}G_{i}-\sum_{i\neq j}\cos G_{i}\cos G_{j})\\
&+&3(t_{\pi }-t_{\sigma })^{4}(\sum_{i}\sin G_{i})^{2}, \end{eqnarray*}
where $G_{i}={\bf a}_{i}{\bf k,}$ and ${\bf a}_{i}$ are the three smallest
lattice vectors. 
At the $\Gamma $ point $v_{{\bf k}}=0,$ and there are two
doubly degenerate states. The bonding pair forms the 2D Fermi surfaces, and
it appears that 
the main effect of the $E_{2g}$ phonons is to lift the degeneracy
at $\Gamma $,  thereby changing the  overall splitting between the two subbands.
This effect does not depend on the sign of the ionic displacement (the
degeneracy is lifted either way) and thus is nonlinear by definition. 
This is 
illustrated in  Fig.\ \ref{fs} by the two Fermi surface plots for 
two opposite $E_{2g}$
phonon patterns, both corresponding to $Q\approx 1$. One can see that
while for the 3D sheets the coupling is mostly linear (changes of the Fermi
surface are opposite), for the 2D cylinders it is mostly quadratic (changes
are the same, cf. the undistorted Fermi surface in Ref.\cite{kortus01}).
Nonlinear EPC is also a likely source of anharmonicity: 
as discussed above, the contribution to the $E_{2g}$ phonon self-energy
from the EPC with the 2D FSs amounts to as much as 20 meV,
as evidenced by the softening of this phonon at
$\Gamma$. A sizeable part of this softening probably
comes from two-phonon processes.
Quartic  anharmonicity of the ion-ion interaction
arises in the fourth order in the linear interaction constant,
but in the second order in the nonlinear one.

In summary, we have presented a first-principles investigation of the
electron-phonon coupling in MgB$_{2}$. Interband anisotropy enhances the
coupling constant from its isotropic dirty-limit value of 
$\lambda _{sc}^{0}=0.77$ to an effective clean-limit value of $\lambda
_{sc}^{eff}=1.01$ for superconductivity. 
With $\omega _{\ln }=56.2$ meV,
this $\lambda _{sc}^{eff}$ is arguably consistent with the measured $T_{c}$
of nearly 40 K.  In the clean limit, we predict two different
superconducting order parameters: a larger one on the 2D FSs and
a smaller one (by approximately one third) on the 3D FSs. 
Since current experiments suggest that
MgB$_{2}$ is indeed in  the clean limit, 
multiple gaps should be observable. 
There are hints of this in both the tunneling and thermodynamic data. 

The $E_{2g}$ phonon mode involving in-plane B motion provides the strongest
coupling. We predict a hardening of  $\sim 12$\%
of this mode at the zone center below $T_{c}$. In
addition, this mode is highly anharmonic and it may also have
significant nonlinear electron-phonon coupling.  The former
likely reduces the linear EPC, while the latter
increases the total EPC.   Further work is needed to
better elucidate the effect of the anharmonicity and nonlinear coupling on the
superconducting properties of this material.


We thank Jim Freericks for helpful discussions. This work was partially supported by the National Science
Foundation under Grant DMR-9973225, and by ONR. AYL acknowledges support
from the ASEE-US Navy Faculty Sabbatical Program.

\begin{table}[tbp]
\caption{Band-decomposition of the electronic density of states at the Fermi
level and in-plane and out-of-plane plasma frequencies. The density of
states is in units of states/Ry/spin/cell, and the plasma frequency is in
eV. }
\label{tabI}%
\begin{tabular}{cccccc}
& total & 1 & 2 & 3 & 4 \\ 
\tableline $N(E_F)$ & 4.83 & 0.66 & 1.38 & 1.26 & 1.52 \\ 
$\omega_{p,xx}$ & 7.21 & 2.91 & 2.95 & 3.05 & 5.04 \\ 
$\omega_{p,zz}$ & 6.87 & 0.44 & 0.52 & 4.62 & 5.06%
\end{tabular}%
\end{table}

\begin{table}[tbp]
\caption{Band-decomposition of the electron-phonon interaction. }
\label{tabII}%
\begin{tabular}{cccccc}
$ij$ & 11 & 12 & 13 & 14 & 22 \\ 
$U_{ij}$~(Ry) & 0.676 & 0.419 & 0.064 & 0.096 & 0.477 \\ 
\tableline 
$ij$ & 23 & 24 & 33 & 34 & 44 \\ 
$U_{ij}$~(Ry) & 0.064 & 0.097 & 0.113 & 0.106 & 0.092%
\end{tabular}%
\end{table}

\begin{table}[tbp]
\caption{Calculated superconducting and transport electron-phonon coupling
parameters in both the isotropic limit and with interband anisotropy.  
The last column contains an average $\lambda_{tr}$ appropriate
for polycrystalline samples. }
\label{tabIII}%
\begin{tabular}{lcccc}
& $\lambda_{sc}$ & $\lambda_{tr,x}$ & $\lambda_{tr,z}$ & $\lambda_{tr,ave}$
\\ 
\tableline
isotropic & 0.77 & 0.70 & 0.46 & 0.58 \\ 
multi-gap& 1.01 & 0.58 & 0.46 & 0.53%
\end{tabular}%
\end{table}


\begin{references}
\bibitem{note3}
In Ref.\ \cite{rubio}, 
where high-quality spectra with a gap three times smaller than the BCS 
value were obtained, a unique procedure was used for sample preparation, 
which may have preferentially oriented individual microcrystals.
\bibitem{isotopenote} For linear coupling, the total isotope
effect of all components must add up to 0.5; nonlinear coupling may change
this total in either direction.
\end{references}
\end{document}